\documentclass[10pt]{article}
\usepackage{amsmath,amsfonts,amssymb}
\usepackage{amscd}
\usepackage{amsthm}
\usepackage{color}
\usepackage{latexsym}
\usepackage{a4wide}
\usepackage{graphicx}
\usepackage{enumerate}
\usepackage{hyperref}
\usepackage{ulem}
\theoremstyle{definition}     \newtheorem{remark}{Remark}
\theoremstyle{plain}          \newtheorem{theorem}{Theorem}
\theoremstyle{definition}     
\theoremstyle{plain}          \newtheorem{lemma}{Lemma}
\theoremstyle{definition}     \newtheorem{corollary}{Corollary}
\theoremstyle{definition}     
\theoremstyle{definition}     \newtheorem{definition}{Definition}
\theoremstyle{plain}          \newtheorem{proposition}{Proposition}
\theoremstyle{plain}          
\theoremstyle{plain}          
\usepackage[top=1.8cm, bottom=1.8cm, left=2.8cm, right=2.8cm]{geometry}
\usepackage{color}
\numberwithin{equation}{section}
\begin{document}
\title{Consolidating Marginalism and Egalitarianism: A New Value for Transferable Utility Games}
\date{}
\author{}
\author{Dhrubajit Choudhury\footnote{Department of Mathematics, Dibrugarh University, Dibrugarh, Assam-786004, India; Email:dhrubajit@gmail.com}~~~Surajit Borkotokey\footnote{Department of Mathematics, Dibrugarh University, Dibrugarh, Assam-786004, India; Email: surajitbor@yahoo.com}~~~Rajnish Kumar\footnote{Economics Group, Queen's Management School, Queen's University Belfast, UK;Email: rajnish.kumar@qub.ac.uk}~~~Sudipta Sarangi\footnote{Department of Economics, Virginia Tech, USA; Email: ssarangi@vt.edu}}

\maketitle
\begin{abstract}
In cooperative games with transferable utilities, the Shapley value is an extreme case of marginalism while the Equal Division rule is an extreme case of egalitarianism. The Shapley value does not assign anything to the non-productive players and the Equal Division rule does not concern itself to the relative efficiency of the players in generating a resource. However, in real life situations neither of them is a good fit for the fair distribution of resources as the society is neither devoid of solidarity nor it can be indifferent to rewarding the relatively more productive players. Thus a trade-off between these two extreme cases has caught attention from many researchers. In this paper, we obtain a new value for cooperative games with transferable utilities that adopts egalitarianism in smaller coalitions on one hand and on the other hand takes care of the players' marginal productivity in sufficiently large coalitions. Our value is identical with the Shapley value on one extreme and the Equal Division rule on the other extreme. We provide four characterizations of the value using variants of standard axioms in the literature. We have also developed a strategic implementation mechanism of our value in sub-game perfect Nash equilibrium.   

\end{abstract}
\section{Introduction}\label{sec:1} 
Cooperative games with transferable utilities or simply TU games describe situations where a finite set of players make binding agreements to generate worths. Their applications in economic allocation problems are wide and varied (see~\cite{tijs,driessen1}).  The underlying assumption is that the players form the grand coalition under such binding agreements. A solution is a rational distribution of the worth of the grand coalition among the players. We call a single point solution a value. Many of the values found in the literature revolve around the notions of egalitarianism and marginalism. Values that combine both these attributes are called solidarity values \cite{radzik,sprumont}. In this paper, we propose a new solidarity value for TU games that focuses on egalitarianism in smaller coalitions and marginalism in larger coalitions. We provide four characterizations of this value on the basis of the coalitional sizes. We also propose a bidding mechanism to implement our value in sub-game perfect equilibrium.

Among all the values found in the literature, the Shapley value~\cite{shapley} is perhaps the most popular one that builds on the notion of marginalism~\cite{young}. It is the expectation of the increase of the transferable utilities of a player when she joins a coalition~\cite{brink2013}, this we call her marginal contributions. Unlike the Shapley value, the Equal Division(ED) rule allocates the worth of the grand coalition equally among all the players and can be considered to be the most egalitarian solution to TU games. Combined with the standard axioms of efficiency, additivity, and symmetry, the Shapley value and the ED are characterized by the null player and the nullifying player property respectively (see \cite{brink2007}). The null player property requires that a completely unproductive player should earn a zero payoff.  The nullifying player property, on the other hand, assigns zero payoffs to players who not only contribute nothing to a coalition but also prevent the production of that coalition. Thus, the two extreme characteristics of marginalism and egalitarianism are realized in the two values by the axioms of the null player and the nullifying player respectively. Different trade-offs between marginalism and egalitarianism are proposed and accordingly, several solidarity values are obtained (see for example \cite{beal,beal2,casajus11,casajus12,casajus14, driessen,joosten,radzik}).  

In our society, egalitarianism is observed in small coalitions while marginalism is observed in large coalitions. An excellent argument for this can be found in Jared Diamond's book `Guns, Germs, and Steel', where he argues that unequal development of regions can be traced through agriculture and the range of animals and species available to domesticate defined the initial development and climate. In most of the states in the northeastern part of India with large tribal populations, this is even visible today. There exist different tribes \footnote{According to the 2011 Census reports, the state of Assam in India has approximately 13\% of the tribal population out of a total population of 30.57 million.} in the hills and plains who have their autonomous councils (constituted based on the sixth schedule to the Constitution of India \footnote{There are all total 9 autonomous councils in the state of Assam, India who enjoy varying degrees of autonomy within the state legislature.}) to look after the welfare of their communities through standard and pre-defined public distribution systems. Both the central and the state governments allocate funds to these autonomous councils based mainly on their marginal productivities to the state exchequer combined with their political and socio-economic credibilities. This allocation to the larger coalition of tribes is governed by marginalism. However, within each such autonomous council, ideally, most of the facilities are community-based and people practice egalitarian distribution and sharing of resources. It is observed that in a coalition of sufficiently small size for example, within a community or a tribe or clan, players are more friendly, homogenous, and socially involved and hence they allow more often the egalitarian sharing of the resources among themselves. However, when more people enter into the coalition making it sufficiently large and heterogeneous, productive players prefer not to share their earnings equally with non-productive or less productive players.

\indent Motivated by the above discussion, in this paper  we first explore whether there exists a value for TU games that embodies both marginalism and egalitarianism depending on the size of the coalitions. Since the difference between the Shapley value and the ED can be attributed to the axioms of the null player and the nullifying player, our starting point here is to define a player who nullifies the worths of all the coalitions of sufficiently small size and becomes non-productive (a null player) in all larger coalitions: call her a  $k$-nullifying null player. A player is a $k$-nullifying null player if her presence in all coalitions of size till $k$ makes them non-productive (zero worth) and beyond size $k$, she does not contribute to any coalition. The value of $k$ determines the point at which the norms switch. The corresponding axiom of $k$-nullifying null property suggests that the proposed value gives zero payoffs to the $k$-nullifying null player. Note that the $k$-nullifying null player is a null player when $k=1$ and a nullifying player when $k=n$. We propose a value: the $k$-SED value, that guarantees egalitarian shares to the non-productive players within small groups e.g., families, communities, tribes depending on the permissive value of size $k$. On the other hand, it also assures that more productive players are not deprived of their marginal productivities insufficiently larger coalitions. The class of values that satisfy efficiency, symmetry, and linearity are called ESL values (see \cite{driessen2003,radzik13}). Most of the values of TU games being ESL values, i.e., characterized by efficiency, symmetry, and linearity differ by a fourth axiom, which is, for example, the null player property for the Shapley value~\cite{shapley}, the nullifying player property for the ED ~\cite{brink2007}, the A-null player property for the solidarity value~\cite{radzik}, the $\delta$-reducing player for the Discounted Shapley value \cite{brink_funaki}, to name a few. The $k$-SED value for each $k$ is an ESL value. Linearity is, however, often considered as a technical condition with little economic implications. Therefore, alternative characterizations of most of these linear values are done using different notions of monotonicity~\cite{casajus14,brink_funaki,yokote17}, the pioneering work with this approach being that of Young~\cite{young}. We, therefore, propose three alternative characterizations avoiding linearity along the lines of these works. Recently, in \cite{beal}, a new class of solidarity values is proposed. The class of $k$-SED values resembles this class of values, however, they differ by the narratives of the problem. We highlight these differences towards the end of the paper. Finally, we propose the implementation mechanism of our value which goes along the line of \cite{beal,beal2}.

\indent The rest of the paper proceeds as follows. In Section~\ref{sec:2} we present the preliminary definitions and results pertaining to the development of the paper. Section~\ref{sec:3} describes a procedure to compute the  $k$-SED value followed by its various characterizations in Section~\ref{sec:4}. In Section~\ref{sec:5}, we compare our model with some of the existing models. Section~\ref{sec:6} describes the implementation mechanism of the class of $k$-SED values and finally, Section~\ref{sec:7} concludes.


\section{Preliminaries}\label{sec:2}

Let $N =\{1,2,...,n\}$ be the player set with $n$ players and $2^N$ denote the power set of $N$. The subsets of $N$ are called coalitions. Denote the size of a coalition $S$ by the corresponding lower case letter $s$. To simplify notation, we write $S\cup i$ for $S\cup \{i\}$ and $S\setminus i$ for $S\setminus \{i\}$ for each $S\subseteq N$ and $i \in N$. A cooperative game or a simple TU game is a pair $(N, v)$ where the function $v : 2^N \rightarrow \mathbb{R}$ is such that $v(\emptyset)=0$.  For each $S \subseteq N$, $v(S)$ denotes the worth of the coalition $S$. If the player set $N$ is fixed, we represent a TU-game by $v$ only. Denote by $v_0$ the \textit{null game}, defined as $v_0(S)=0$ for all $S \subseteq N$. A TU game is \textit{zero-monotonic} if, for each $i\in  N$ and each coalition $S \ni i$, it holds that $v(S) - v(S\setminus i) \geq v(i)$. The class of all TU games over the player set $N$ is denoted by $G(N)$. Recall that the standard assumption of TU games is that the grand coalition is eventually formed. A solution of an $n$-player TU-game is an $n$-dimensional payoff vector $\mathbf{x}\in \mathbb{R}^n$  giving a payoff $x_i\in \mathbb{R}$ to every player $i\in N$. A value on $G(N)$ is a function $\Phi$ that assigns a payoff vector $\Phi(v)\in \mathbb{R}^n$ to each $v\in G(N)$ for a fixed player set $N$. The class $G(N)$ of all TU games with player set $N$ forms a vector space of dimension $2^n-1$ under the standard addition and scalar multiplications of set functions. For every coalition $S\subseteq N$ with $S\neq \emptyset$, the games $e_S : 2^N \rightarrow \mathbb{R}$ and $u_S: 2^N \rightarrow \mathbb{R}$ given by, 
 \begin{equation}\label{eq:basis1}
 e_{S}(T)= \left\{
 \begin{array}{cc}
 1,~~~\mbox{if} ~T= S &  \\ 
 0, ~~~\mbox{otherwise,} &\\ 
 \end{array}
 \right.
 \end{equation} 
 
 \begin{equation}\label{eq:basis2}
 u_{S}(T)= \left\{
 \begin{array}{cc}
 1,~~~\mbox{if} ~ S\subseteq T &  \\ 
 0, ~~~\mbox{otherwise,} &\\ 
 \end{array}
 \right.
 \end{equation}
are standard bases for the class $G(N)$ of TU games with player set $N$ called the identity games and the unanimity games respectively.
For every game $v\in G(N)$, we can write $v=\sum_{S\not =\emptyset} v(S)e_S$ and $v=\sum_{S\not =\emptyset} \Delta_S(v)u_S$ where $\Delta_S(v)=\sum_{T\subseteq S} (-1)^{s-t}v(T)$. 
The marginal contribution of player $i$ to coalition $S$ is formally written as, 
 \begin{equation}
 m_i^v(S)=v(S\cup i)-v(S).
 \end{equation}
Suppose that the grand coalition $N$ is formed in such a way that the players enter the coalition one by one. Such entry can be attributed to a permutation $\pi: N\rightarrow N$ of the players. We denote the collection of all permutations by $\Pi(N)$. For every $\pi\in \Pi(N)$, we denote by $P(\pi , i)=\{j\in N: \pi(j)<\pi(i)\}$ the set of players that enter before the player $i$ in the permutation $\pi$. The Shapley value \cite{shapley} denoted by $\Phi^{Sh}$ assigns to every player her expected marginal contribution (to the coalition of players that enters before her), given that every permutation of entrance $\pi$ has equal probability of occurrence namely, $\dfrac{1}{n!}$. \\
Therefore the Shapley value is given by
 
 \begin{equation}
 \Phi^{Sh}_i(v)= \frac{1}{n!}\sum_{\pi\in \Pi(N)} m_i^v(P(i,\pi)),
 \end{equation} 
which after simplifications becomes, 
 
 \begin{equation}\label{eq:shapley}
 \Phi^{Sh}_i(v)= \sum_{S\subseteq N\setminus i} \frac{{s!}\left({n-s - 1}%
 	\right)!}{n!}m_i^v(S).
 \end{equation} 

\vspace{4pt}

\indent The Equal Division (ED) rule is a solution $\Phi^{ED}:G(N)\rightarrow \mathbb{R}^n $ that distributes the worth $v(N)$ of the grand coalition equally among all players in any game, i.e., 
\begin{equation}\label{eq:equaldivision}
\Phi^{ED}_i(v)=\dfrac{v(N)}{n}.
\end{equation}
It follows from (\ref{eq:shapley}) and (\ref{eq:equaldivision}) that both the Shapley value and the Equal Division rule can be expressed in a unified manner as follows:
\begin{equation}
\Phi_i(v) = \sum_{S\subseteq N} \frac{{s!}\left({n-s-1}
 	\right)!}{n!}C_i^v(S),
\end{equation}
where the quantity $C_i^v(S)\in \mathbb{R}$ be such that $C_i^v(S)  = m_i^v(S)\;\forall\;S\subseteq N,$ when $\Phi = \Phi^{Sh}$ and $C_i^v(S) = e_{N}(S) v(S)\;\forall\;S\subseteq N,$ when $\Phi=\Phi^{ED}$. 

Call $C_i^v(S)$ the coalitional contribution of player $i$ in $S$ with respect to $v\in G(N)$. Thus, under this new notation, both the Shapley value and the ED assign to each player her expected coalitional contribution $C_i^v(S)$ where $C_i^v(S) = m_i^v(S)$ in case of the Shapley value and $C_i^v(S) = 0$ for $S\subsetneq N$ and;  $C_i^v(S) = v(N)$ for $S=N$, in case of the Equal Division rule. 


Various axiomatizations of the Shapley value and the Equal Division rule can be found in the literature (see ~\cite{chun,shapley,brink2,brink2007,young}). In the following, we list some of the important axioms that characterize these two values and also are relevant to the present paper. Prior to that we define the following:
\begin{definition}
A player $i\in N$ is a null player in $v$ if $m_i^v(S)=0$ for every coalition $S\subseteq N$. 
\end{definition}
\begin{definition}
A player $i\in N$ is a nullifying player in $v$ if $v(S)=0$ for every coalition $S$  with $i\in S$.  
\end{definition}
\begin{definition}
Two players $i,j \in N$ are called symmetric with respect to the game $v$ if for all $S \subseteq N\setminus \{i,j\}$, $$v(S\cup i) = v(S \cup j).$$
\end{definition}
We list the axioms for a value $\Phi : G(N) \rightarrow\mathbb{R}^n$  as follows:
\begin{enumerate}[\textbf{Axiom} 1.]
	\item Efficiency (Eff): $\Phi$ is efficient if for any $v\in G(N)$ : 	
	$\sum_{i\in N} \Phi_i(v) = v(N).$
	
	\item Null Player Property (NP): For every game $v\in G(N)$ and every null player $i \in N$ in $v$, we have $\Phi_i(v) = 0$ . 	
	\item Nullifying Player Property ({NPP}): For every game $v\in G(N)$ and every nullifying player $i \in N$ in $v$, $\Phi_i(v) = 0$. 
	\item Symmetry (Sym): For every pair of symmetric players $i,j\in N$ with respect to the game $v \in G(N)$, we have $\Phi_i(v)= \Phi_j(v)$. 
	\item Linearity (Lin): For all games $u, w \in G(N)$, every pair of $\gamma, \eta\in \mathbb{R}$, and every player $i\in N$:  
	\begin{equation}\label{eq:lin}
	\Phi_i(\gamma u+\eta w)= \gamma \Phi_i(u)+ \eta \Phi_i(w);
	\end{equation}
	$\Phi$ is additive (ADD) if in particular (\ref{eq:lin}) holds for $\gamma= \eta=1$.
	\item Strong monotonicity (SMon): $\Phi_i(v) \geq \Phi_i(w)$ for every pair of games $v,w\in G(N)$ and player $i\in N$ such that $m_i^v(S)\geq m_i^w(S)$ for all $S\subseteq N\setminus i$. 
	\item Coalitional strategic equivalence (CSE): For every pair of games $v,w\in G(N)$, a value $\Phi$ satisfies $\Phi_i(v+w)=\Phi_i(v)$ whenever $i$ is a null player in $w$. 
	\item Fairness\footnote{In \cite{casajus11} this axiom is termed as the \textit{van den Brink fairness} since it was introduced by van den Brink in \cite{brink2}.} (F): For any two symmetric players $i,j\in N$ in $w\in G(N)$, it holds that 
	\begin{equation*}
	\Phi_i(v+w)-\Phi_i(v)=\Phi_j(v+w)-\Phi_j(v),  \qquad\forall v\in G(N).
	\end{equation*} 
	\item Desirability (D): A value satisfies Desirability if for all $v\in G(N)$ and all $i,j\in N$,  $v(S\cup i)\geq v(S\cup j)$ for all $S\subseteq N\setminus \{i,j\}$, then $\Phi_i(v)\geq \Phi_j(v)$. 
	\item Differential marginality (DM): For all $v, w \in G(N)$, $i,j \in N$, $v(S \cup i) - v(S \cup j) = w(S\cup i)- w(S \cup j)$ for all $S \subseteq N \setminus \{i,j\}$ implies that $\Phi_i(v)- \Phi_j(v) = \Phi_i(w)- \Phi_j(w)$.
\end{enumerate} 
The most standard characterization of the Shapley value requires Eff, Sym, Lin, and NP. The ED, on the other hand, has been characterized using Eff, Sym, Lin, and NPP (see~\cite{brink2007}). An alternative characterization of the Shapley value is due to \cite{young} that uses Eff, Sym, and SMon. Chun~\cite{chun} characterizes the Shapley value using Eff, Sym, and CSE. Finally, van den Brink \cite{brink2} characterizes the Shapley value by the axioms of Eff, NP, and F. Recall from Section \ref{sec:1} that a value that satisfies Eff, Sym, and Lin is called an ESL value \cite{driessen2003}. We mention the following proposition from \cite{radzik} for later reference. 
 
\indent Recall from Section~\ref{sec:1} that a value that satisfies Eff, Sym and Lin is called an ESL value \cite{driessen2003}. We mention the following proposition from \cite{ESLvalue} for later reference.
\begin{proposition}\label{prop:esl1}(Proposition 2 in \cite{ESLvalue}, p. 184) \rm  
A value $\Phi$ on $G(N)$ is an ESL-value if and only if there exists a unique collection of real constants $B^{\Phi}=(b^{\Phi}_s: s\in \{0,1,2,...,n\})$ with $b^{\Phi}_n=1$ and $b^{\Phi}_0=0$ such that for every game $v\in G(N)$,
\begin{equation}\label{eq:esl}
\Phi_i(v)=\sum_{S\subseteq N\setminus i} \dfrac{s!(n-s-1)!}{n!}\bigg\{b^{\Phi}_{s+1}v(S\cup i)-b^{\Phi}_sv(S)\bigg\}.
\end{equation}
That is
\begin{equation}\label{Eq:gen}
\Phi_i(v)=\Phi^{Sh}_i(B^{\Phi} v)
\end{equation} 
where $(B^\Phi v)(S)=b^\Phi_s v(S)$ for each coalition of size $s$. This value $\Phi$ is denoted by $\Phi^{ESL}$.
\end{proposition}

\section{The $k$-SED value}\label{sec:3}
We now introduce our new value for TU Cooperative games, which we term the $k$-SED value. We follow an approach similar to Shapley's~\cite{shapley} approach where the players are allowed to enter a coalition following a particular permutation assuming that all possible permutations of entrance have equal probabilities.
Our value is based on the assumption that given a coalition $S$ of sufficiently small size, each player $i$ agrees to the egalitarian distribution of its worth $v(S)$, namely $\frac{v(S)}{s}$: let us call it the egalitarian coalitional contribution. However, when the size of the coalitions is sufficiently large, the coalitional contributions become marginal contributions $m_i^S(v)$ and are no longer egalitarian.  

\indent Let $k$ be the maximum size of the coalitions in which each player enjoys egalitarian coalitional contributions. Recall that the players enter the coalition one by one. Define $P^k(\pi)=\{j\in N|\pi(j)\leq k\}$. Then $P^k(\pi)$ represents the set of first $k$  players who enter the game under the permutation $\pi$. If $\pi(i)\leq k$ then $P(\pi,i)\cup i \subseteq P^k(\pi)$. Thus, following our assumption the coalitional contribution $C_{i}^{v}(P(\pi,i))$ of player $i$ is $\dfrac{v(P^k(\pi))}{k}$ when it enters into $P(\pi,i)$ and when $\pi(i)\leq k$. On the other hand, when player $i$ enters the coalition $P(\pi,i)$ with $\pi(i)>k$, then $C_{i}^{v}(P(\pi,i))= v(P(\pi,i)\cup i)-v(P(\pi,i))$ in the permutation $\pi$.  Thus,  the coalitional contribution $C_{i}^{v}(P(\pi,i))$ of player $i$ in a game $v$ in forming the grand coalition $N$ following the permutation $\pi$ is given by
\begin{equation}\label{eq:payoff}
C_{i}^{v}(P(\pi,i)) = \left\{
\begin{array}{cc}
\dfrac{v\bigg(P^k(\pi)\bigg)}{k},~~~\mbox{if} ~ \pi(i)\leq k &  \\ 
v(P(\pi,i)\cup i)-v(P(\pi,i)), ~~~\mbox{if} ~\pi(i)>k. &\\ 
\end{array}
\right.
\end{equation}
\vspace{4pt} 

\noindent The $k$-SED value denoted by $\Phi^{k-\rm{SED}}: G(N)\rightarrow \mathbb{R}^n$ is the value that assigns to every player $i\in N$, its expected coalitional contribution given by ~(\ref{eq:payoff}), i.e.,
\begin{align*}\label{eq: exp1}
\Phi_i^{k-SED}(v)&=\dfrac{1}{n!}\sum_{\pi\in \Pi(N)}C_i^{v}(P(\pi,i))\\
&=\dfrac{1}{n!}\sum_{\pi\in \Pi(N):\pi(i)\leq k}\dfrac{v(P^k(\pi))}{k}+\dfrac{1}{n!}\sum_{\pi\in \Pi(N):\pi(i)> k}\bigg\{v(P(\pi,i)\cup i)- v(P(\pi,i))\bigg\}\\
&=\dfrac{1}{n!}\sum_{\pi\in \Pi(N):\pi(i)= k}k\dfrac{v(P^k(\pi))}{k}+\dfrac{1}{n!}\sum_{\pi\in \Pi(N):\pi(i)> k}\bigg\{v(P(\pi,i)\cup i)- v(P(\pi,i))\bigg\}\\
&=\dfrac{1}{n!}\sum_{\pi\in \Pi(N):|P(\pi,i)|= k-1}v(P(\pi,i)\cup i)+\dfrac{1}{n!}\sum_{\pi\in \Pi(N):|P(\pi,i)|\geq k}\bigg\{v(P(\pi,i)\cup i)- v(P(\pi,i))\bigg\}
\end{align*}
After simplification, it can be re-written as follows:
\begin{equation}\label{eq:exp2}
\Phi_i^{k-SED}(v)=\sum_{\substack {S\subseteq N\;:\;i \not \in S \\  s= k-1}}\dfrac{(n-k)!(k-1)!}{n!}v(S\cup i)+\sum_{\substack {S\subseteq N\;:\;i\not \in S \\  s\geq k}} \dfrac{(n-s-1)!s!}{n!}\bigg\{v(S\cup i)-v(S)\bigg\}.
\end{equation}

\begin{remark}\label{rem:1} 
\begin{enumerate}[~~~(a)]
\item The $k$-SED value coincides with the Shapley value for $k=1$ and the ED for $k=n$. 
\item For each $v\in G(N)$, $\Phi^{k-SED}(v)= \Phi^{Sh}(B^k v)$ where $B^k=(b_s:s\in \{0,1,2,...,n\})$ such that 	$b_0=0$, $b_s=0$ for $s<k$, $b_s=1$ for $s\geq k$.\label{it:d}
		\item In view of (\ref{it:d}) above, for each $v\in G(N)$, define $\bar v \in G(N)$ as follows.
		\begin{equation}\label{eq:newgame}
		\bar v(S) = \left\{
		\begin{array}{cc}
		0 ,~~~\mbox{if} ~ s< k &  \\ 
		v(S), ~~~\mbox{if} ~s\geq k &\\ 
		\end{array}
		\right.
		\end{equation} 
		Then $\Phi^{k-SED}_i(v)=\Phi^{Sh}_i(\bar v)$.  
\end{enumerate}
\end{remark}

\vspace{4pt}
 
\section{Characterization} \label{sec:4} 
In this section we characterize $\Phi^{k-\rm{SED}}$ using four sets of axioms and explore their relationships with the various characterizations of the Shapley value mentioned in Section~\ref{sec:2}.
\subsection{The $k$-nullifying null player} \label{subsection:4.1}
As we have defined in Section~\ref{sec:1}, we call a player that nullifies the contributions of the small coalitions and becomes non-productive in sufficiently large coalitions a $k$-nullifying null player. Formally, we have
\begin{definition}
Let $k \in \{1,2,...,n\}$ be given. Player $i\in N$ is a $k$-nullifying null player if $v(S\cup i)=0$ for $S\subseteq N\setminus i$ with $s<k$ and $v(S\cup i)=v(S)$ for $S\subseteq N\setminus i$ with $s\geq k$.   
\end{definition}
\begin{enumerate}[\textbf{Axiom} 11.]
\item $k$-nullifying null player property: ($k$-{NNPP}): A value $\Phi : G(N) \rightarrow\mathbb{R}^n$ satisfies $k$-nullifying null player property if for every $v\in G(N)$ it holds that $\Phi_i(v)=0$ for every $k$-nullifying null player $i\in N$. 
\end{enumerate}
Note that $k$-{NNPP} requires that a player that annihilates the contributions of the small coalitions and becomes non-productive in sufficiently large coalitions should be rewarded zero payoff. Replacing the NP with the $k$-{NNPP} property in the characterization of the Shapley value we obtain the characterization of the $k$-SED value. In the following, we provide the first characterization theorem where we show that the $k$-SED value is efficient, linear, symmetric and it gives zero payoffs to those players who make the small coalitions non-productive and become themselves non-productive in sufficiently large coalitions. Unless specified, we keep $k$ fixed here.

\indent Let us introduce two subspaces of $G(N)$ as follows:
$$G_{<k}(N)=\{v\in G(N):v(S)=0 \,\,\text{for all } s\geq k \}\;\;\textrm{and}\;\; G_{\geq k}(N)=\{v\in G(N):v(S)=0 \,\,\text{for all } s< k \}.$$
Then using the standard notation for the direct sum of linear spaces, we get $G(N)=G_{<k}(N)\oplus G_{\geq k}(N)$. It follows that every game $v\in G(N)$ can be written as $v=v_{<k}+ v_{\geq k}$ where $v_{<k}\in G_{<k}(N)$, $v_{\geq k}\in G_{\geq k}(N)$ such that $v_{<k}(S)=v(S)$ for all $s<k$,  $v_{<k}(S)=0$ for all $s\geq k$ and $v_{\geq k}(S)=v(S)$ for all $s\geq k$,  $v_{\geq k}(S)=0$ for all $s< k$.

Next, we define a basis for the class $G(N)$ which will be useful in showing the uniqueness of the $k$-SED value at a later stage. Observe that $\{e_S:S\subseteq N, s<k\}$ given by (\ref{eq:basis1}) is a basis of $G_{<k}(N)$ and $\{u_S:S\subseteq N, s\geq k\}$ given by (\ref{eq:basis2}) is a basis of $G_{\geq k}(N)$. Since $G(N)=G_{<k}(N) \oplus G_{\geq k}(N)$, therefore, the set $W=\{w_S:S\subseteq N, S\neq \emptyset\}$ where each $w_S \in G(N)$ is defined by 
\begin{equation}\label{eq:basis3}
	w_{S}(T)= \left\{
	\begin{array}{cc}
		e_S(T),~~~\mbox{if} ~ s<k &  \\ 
		u_S(T), ~~~\mbox{if} ~s\geq k &\\ 
	\end{array}
	\right.
\end{equation}
 is a basis for $G(N)$. Observe that every game $v\in G(N)$ can be written as $v=\sum_{S\neq \emptyset } \lambda^k_S(v) w_S$ where $ \lambda^k_S(v)=\sum_{T\subset S: t\geq k}(-1)^{s-t}v(T)$ for $s\geq k$ and $ \lambda^k_S(v)=v(S)$ for $s<k$. In the following, we give the first characterization theorem of the $k$-SED value.
\begin{theorem}\label{them:1}\rm      
The following statements are equivalent:
\begin{enumerate}[~~\rm(1)]
		\item $\Phi$ satisfies { Eff, Sym, Lin} and $k$-\rm{NNPP}.
		\item $\Phi$ is given by $\Phi_i(v)=\sum_{S: i\in S}\dfrac{\lambda^k_S(v)}{s}$
		\item 	$\Phi(v)=\Phi^{\rm{ED}}(v_{<k})+\Phi^{\rm{Sh}}(v_{k\geq k})$. 
		\item $\Phi=\Phi^{k-\rm{SED}}$ 
\end{enumerate}
\end{theorem}
\begin{proof}
By Lin, $\Phi$ is unique if it is unique on a basis. By Eff, Sym, and $k$-NNPP, $\Phi$ is unique on $W=\{w_S:S\subseteq N, S\neq \emptyset\}$. Moreover, $\Phi_i(v)=\sum_{S: i\in S}\dfrac{\lambda^k_S(v)}{s}\;\;$
 clearly satisfies Eff, Sym, Lin and $k$-NNPP. This establishes (a)$\Leftrightarrow$(b). It is obvious that 
(c)$\Leftrightarrow$(d). Finally, $\Phi(v)=\Phi^{\rm{ED}}(v_{<k})+\Phi^{\rm{Sh}}(v_{k\geq k})$ satisfies Eff, Sym, Lin and $k$-NNPP. This completes the proof.
\end{proof}
\begin{remark}\label{rem:2}
In view of theorem~\ref{them:1}, $\Phi^{k-SED}$ is an ESL value. Therefore, by proposition~\ref{prop:esl1}, the formula for $\Phi^{k-SED}$ given by (\ref{eq:exp2}) has the equivalent form given by (\ref{eq:esl}) with $b_j^\Phi=0$ for all $j<k$ and $b_j^\Phi = 1$ for all $j\geq k$. 
\end{remark}
\subsubsection*{Logical Independence of the axioms in Theorem \ref{them:1}}
\indent In the following, we show the logical independence of the axioms in Theorem \ref{them:1}.\\
\noindent\textbf{Dropping Eff}: The value $\Phi^1:G(N)\rightarrow \mathbb{R}^n$ given by $\Phi^1_i(v)=\dfrac{1}{2^{n-1}}\sum_{S \subseteq N: s\geq k }\bigg\{v(S\cup i)-v(S)\bigg\}$ satisfies {Sym, Lin} and  $k$-{NNPP}  but does not satisfy Eff.\\
\noindent\textbf{Dropping Lin}: Define a new value  $\Phi^2:G(N)\rightarrow \mathbb{R}^n$ given by $\Phi^2_i(v)=\dfrac{\Phi^1_i(v)}{\sum_{j\in N}\Phi^1_j(v)}v(N)$ for all $i\in N$ if $\sum_{j\in N}\Phi^1_j(v)\neq 0$. Then $\Phi^2$ satisfies { Sym, Eff} and  $k$-{NNPP} but does not satisfy Lin.\\
\noindent\textbf{Dropping $k$-{NNPP}}: The value $\Phi^3:G(N)\rightarrow \mathbb{R}^n$ given by $\Phi^2_i(v)=\dfrac{v(N)}{n}$ satisfies { Sym, Lin, Eff}  but does not satisfy $k$-{NNPP} for $k<n$.\\     
\noindent\textbf{Dropping Sym}: Consider the basis $W=\{w_S:S\neq \emptyset\}$ for $G(N)$. Each $i\not \in S$ is a $k$-nullifying null player for the game $w_S$. Let $\pi:S \rightarrow S$ be a permutation on $S$. Define $r = \min\{\pi(j)|j\in S\}$. Define a value $\Phi^4:G(N)\rightarrow \mathbb{R}^n$ given by $\Phi^4_r(w_S)=w_S(N)$ and $\Phi^4_j(w_S)=0$ for all $j\in N\setminus r$.  
Then $\Phi^4$ satisfies {Lin, Eff} and $k$-{NNPP} but does not satisfy Sym.


\subsection{The coalitional $k$-strategic equivalence}\label{subsection:4.2}
The Shapley value is characterized in ~\cite{brink2007} with the axioms {Eff, Sym} and {CSE}. In~\cite{chun}, another characterization is done using the same set of axioms, but the definition of CSE in \cite{chun} differs slightly from \cite{brink2007}. In \cite{brink2007}, it is shown that these two definitions are equivalent. Therefore, here we use the definition given in \cite{brink2007}. Note that CSE combines {ADD} and {NP} and states that the payoff of a player from any game does not change when another game in which she is a null player is added to it. Here we show that replacing the null player with a $k$-nullifying null player also does not change the payoff of a player if we add a game in which this player is a $k$-nullifying null player. Note that, in particular, when $k=n$, the $n$-nullifying null player is a nullifying player and the proposed property implies that the payoff of a player from any game does not change if another game is added to it in which she is a nullifying player. Thus we have the following:

\begin{enumerate}[\textbf{Axiom} 12.]
	\item Coalitional $k$-strategic equivalence($k$-CSE): A value $\Phi : G(N) \rightarrow\mathbb{R}^n$ satisfies Coalitional $k$-strategic equivalence if for every pair of games $v, w\in G(N)$ it holds that $\Phi_i(v+w)=\Phi_i(v)$ whenever $i$ is a $k$-nullifying null player in $w$. 
\end{enumerate}
\begin{lemma}\label{lemma:4}\rm
If a solution $\Phi:G(N)\rightarrow \mathbb{R}^n$ satisfies Lin and $k$-{NNPP} then $\Phi$ satisfies $k$-CSE. But the converse is not true.  
\end{lemma}
\begin{proof}
By Lin, $\Phi_i(v+w)=\Phi_i(v)+\Phi_i(w)$. Now if $i$ is a $k$-nullifying null player in $w$ then 
$\Phi_i(w)=0$ by $k$-{NNPP}. 
Therefore $\Phi_i(v+w)=\Phi_i(v)$ and hence $\Phi$ satisfies $k$-CSE.  
 
A solution that satisfies $k$-CSE need not satisfy Lin and $k$-{NNPP}. This can be seen from the function $\Phi:G(N)\rightarrow \mathbb{R}^n$ given by $\Phi_1(v)=\Phi_1^{k-\rm{SED}}(v)+2$ and $\Phi_i(v)=\Phi_i^{k-\rm{SED}}(v)-\dfrac{2}{n-1}$ for all $i\in N\setminus 1$. Since $\Phi_i^{k-\rm{SED}}(v+w)=\Phi_i^{k-\rm{SED}}(v)+\Phi_i^{k-\rm{SED}}(w)$ and $\Phi_i^{k-\rm{SED}}(w)=0$ for $k$-nullifying null player $i$ in $w$, therefore $\Phi_i(v+w)=\Phi_i(v)$. Thus $\Phi$ satisfies $k$-CSE but it neither implies Lin nor $k$-{NNPP}. 
\end{proof}
Lemma \ref{lemma:4} implies that the axiom $k$-CSE is weaker than Lin and $k$-{NNPP}. In the following, a characterization of the $k$-SED value is presented based on $k$-CSE.
\begin{theorem}\label{them:2}\rm
A value $\Phi:G(N)\rightarrow \mathbb{R}^n$ satisfies Eff, Sym and $k$-CSE if and only if $\Phi = \Phi^{k-\rm{SED}}$. 
\end{theorem}
\begin{proof}
It is easy to check that $\Phi^{k-\rm{SED}}$ satisfies Eff, Sym and $k$-CSE. 
We will prove the uniqueness by induction on $d(v)=|\{T\subseteq N:v(T)\neq 0\}|$\footnote{Our procedure follows a similar procudure discussed in \cite{brink2007} where the uniqueness for the Equal Division rule is shown by induction on the number of coalitions with non zero dividend.}. If $d(v)=0$ then $v(S)=0$ for all $S\subseteq N$. By Eff and Sym, $\Phi_i(v)=0=\Phi_i^{k-\rm{SED}}(v)$ for all $i\in N$.  Assume that $\Phi_i(w)=\Phi_i^{k-\rm{SED}}(w)$ for all $d(w)<d(v)$. Let $H(v)=\{i\in N:v(S)=0~ \mbox{for all}~S\subseteq N\setminus i\}$. Then for every $i\in N\setminus H(v)$, there exists an $S\subseteq N\setminus i$ such that $v(S)\neq 0$. Since $(v-v(S)e_S)(T)=v(T)$ for $S\neq T$ and $(v-v(S)e_S)(T)=0$ for $S=T$, we have $d(v-v(S)e_S)=d(v)-1$, i.e., $d(v-v(S)e_S)<d(v)$. By induction hypothesis, $\Phi_i(v-v(S)e_S)=\Phi^{k-\rm{SED}}_i(v-v(S)e_S)$ for $i\in N\setminus H(v)$ and $S\subseteq N\setminus i$. Since $\Phi_i(v(S)e_S)=0$ and $i$ is a $k$ null-nullifying player in $v(S)e_S$ therefore $\Phi_i(v)=\Phi_i(v-v(S)b_S)$ by $k$-CSE. Since $\Phi^{k-\rm{SED}}_i(v)=\Phi^{k-\rm{SED}}_i(v-v(S)e_S)$ for $S\subseteq N\setminus i$, therefore, $\Phi_i(v)=\Phi^{k-\rm{SED}}_i(v)$ for $i\in N\setminus H(v)$. With Sym and Eff, it follows that $\Phi_i(v)=\dfrac{v(N)-\sum_{j\in N-\setminus H(v)}\Phi^{k-\rm{SED}}_j(v)}{|H(v)|}=\Phi^{k-\rm{SED}}_i(v)$ for $i\in H(v)$. Therefore $\Phi=\Phi^{k-\rm{SED}}$ by induction hypothesis.  
\end{proof} 
\indent The following corollary is an immediate consequence of Theorem~\ref{them:2}.
\begin{corollary}\label{cor:1}\rm
A value $\Phi:G(N)\rightarrow \mathbb{R}^n$ satisfies Eff, Sym and $n$-CSE if and only if $\Phi = \Phi^{ED}$.
\end{corollary}
\subsubsection*{Logical Independence of the axioms of Theorem~\ref{them:2}}
\noindent\textbf{Dropping Eff}: The value $\Psi^5:G(N)\rightarrow \mathbb{R}^n$ given by 
$$\Psi^5_i(v)=\dfrac{1}{2^{n-1}}\sum_{S \subseteq N: s\geq k }\bigg\{v(S\cup i)-v(S)\bigg\}$$ 
satisfies Sym and $k$-CSE but does not satisfy Eff.\\
\noindent\textbf{Dropping Sym}: Define a value $\Psi^6:G(N)\rightarrow \mathbb{R}^n$ given by $\Psi^6_1(v)=\Phi_1^{k-\rm{SED}}(v)+2$ and $\Psi^6_i(v)=\Phi_i^{k-\rm{SED}}(v)-\dfrac{2}{n-1}$ for all $i\in N\setminus 1$. Clearly $\Psi^6$ satisfies Eff and  $k$-CSE. But $\Psi^6$ does not satisfy Sym. \\  
\noindent\textbf{Dropping $k$-CSE}: Let the value $\Psi^7:G(N)\rightarrow \mathbb{R}^n$ be as defined in (b). Define a new value  $\Psi^7:G(N)\rightarrow \mathbb{R}^n$ given by $\Psi^7_i(v)=\dfrac{\Psi^6_i(v)}{\sum_{j\in N}\Psi^6_j(v)}v(N)$ for all $i\in N$ if $\sum_{j\in N}\Psi^6_j(v)\neq 0$. Then $\Psi^7$ satisfies Sym and Eff but does not satisfy $k$-CSE. 
\subsection{The $k$-partial monotonicity}\label{subsection:4.3}
In~\cite{young}, the elegant notions of Marginality (M) and  Strong Monotonicity(SMon) are introduced in the characterization of the Shapley value. The axiom M states that if a player's marginal contributions are identical in two games then her payoffs from these two games should also be equal. The SMon axiom states that between any two games, a player gets higher payoff from the one in which her marginal contributions are all greater. The axiom of coalitional monotonicity due to van den Brink~\cite{brink2007} states that for each pair of games $v, w \in G(N)$, if $v(S)\geq w(S)$ for all $S\subseteq N$, then $\Phi_i(v)\geq \Phi_i(w)$ for each $i \in N$. Now, we introduce the axiom of $k$-partial monotonicity that weakly combines these two axioms on the basis of the size of coalitions determined by $k$. Note that a similar axiom can also be introduced replacing the notion of monotonicity by marginality. 
\begin{enumerate}[\textbf{Axiom} 13.]
	\item $k$-Partial Monotonicity ($k$-PMon): Given two games $v, w \in G(N)$, a value $\Phi : G(N) \rightarrow\mathbb{R}^n$ satisfies $k$-partial monotonicity if for each $i \in N$, $\Phi_i(v)\geq \Phi_i(w)$  whenever either of the following holds:
	\begin{enumerate}[~~(i)]
	\item $m_i^v(S)\geq m_i^w(S)$ for all $S\subset N$ such that $s\geq k$ with $i \not \in S$.
	\item $v(S)\geq w(S)$ for all $S \subset N$ such that $s< k$ with $i \in S$.
	\end{enumerate} 
\end{enumerate}

\begin{enumerate}[\textbf{Axiom} 14.]
	\item $k$-Partial Marginality ($k$-PM): Given two games $v, w \in G(N)$, a value $\Phi : G(N) \rightarrow\mathbb{R}^n$ satisfies $k$-partial marginality if for each $i \in N$, $\Phi_i(v)= \Phi_i(w)$  whenever either of the following holds:
	\begin{enumerate}[~~(i)]
	\item $m_i^v(S)= m_i^w(S)$ for all $S\subset N$ such that $s\geq k$ with $i \not \in S$.
	\item $v(S)= w(S)$ for all $S \subset N$ such that $s< k$ with $i \in S$.
	\end{enumerate} 
\end{enumerate}

It can be easily shown that the $k$-SED value satisfies $k$-PMon and $k$-PM.

\begin{remark}\label{rem:3}
Following Proposition 3  of \cite{casajus11} (p. 169) it can be easily shown that $k$-PMon and $k$-PM are equivalent and therefore, both imply $k$-CSE. Thus, in view of Theorem~\ref{them:2}, we have the following theorem as its corollary.
\end{remark}
\begin{theorem}\rm
A value $\Phi:G(N)\rightarrow \mathbb{R}^n$ is equal to the $k$-SED value if and only if it satisfies Eff, Sym and $k$-PMon (or $k$-PM). 
\end{theorem}

\subsection{The fairness axiom}\label{subsection:4.4}
In \cite{brink2}, the Shapley value is characterized using the axioms of Eff, NP and F. We propose to replace NP by $k$-NNPP and obtain a characterization of the $k$-SED value along the same line. The following lemma due to \cite{brink2} is useful for our characterization.
\begin{lemma}[\cite{brink2}, Proposition 2.4(i), pg 311]\label{lemma:5}\rm
If a value $\Phi:G(N)\rightarrow \mathbb{R}^n$ satisfies Sym and Lin then $\Phi$ also satisfies F. But the converse is not true. 
\end{lemma}
Similar to Proposition 2.4(ii) in \cite{brink2}, we have the following Lemma.
\begin{lemma}\label{lemma:6}\rm
If a value $\Phi:G(N)\rightarrow \mathbb{R}^n$ satisfies F and $k$-{NNPP} then $\Phi$ satisfies Sym. 
\end{lemma}
\begin{proof}
Suppose that $\Phi$ satisfies F and $k$-{NNPP}. For the null game $v_0\in G(N)$ given by $ v_0(S)=0$ for all $S\subseteq N$, each player $i\in N$ is a $k$-nullifying null player. Therefore $\Phi_i(v_0)=0$ by $k$-{NNPP}. Suppose that $i,j\in N$ are two symmetric players in $v\in G(N)$. By F, $\Phi_i(v_0+v)-\Phi_i(v_0)=\Phi_j(v_0+v)-\Phi_j(v_0)$. Since $(v_0+v)=v$ therefore $\Phi_i(v)=\Phi_j(v)$ and hence $\Phi$ satisfies Sym.
\end{proof} 
\indent It follows from Lemma~\ref{lemma:6} that F is not equivalent to Lin and Sym. Thus we have the following proposition.    
\begin{proposition}\label{lemma:7}\rm
If $\Phi:G(N)\rightarrow \mathbb{R}^n$ satisfies Eff, $k$-{NNPP} and F then,
$\Phi_i(v_0)=0$ for all $i\in N$ where $v_0$ is the null game in $G(N)$ and
\begin{equation*}
    \Phi_i(\lambda w_S)= \left\{
     \begin{array}{cc}
		0,~~~\mbox{if} ~ i\not \in S &  \\ 
		\dfrac{\lambda}{s}, ~~~\mbox{if} ~i\in S. &\\ 
		\end{array}
		\right.
\end{equation*}		
\end{proposition}
\begin{proof}
Suppose $\Phi:G(N)\rightarrow \mathbb{R}^n$ satisfies Eff, the $k$-{NNPP} and F. Then $\Phi$ satisfies Sym by Lemma \ref{lemma:6}. Now the result follows immediately from Eff, $k$-{NNPP} and Sym.
\end{proof}
\indent Next, we introduce some mathematical preliminaries to prove our next characterization. For every $v\in G(N)$, we write $v$ as $v=\sum_{S\subseteq N: S\neq \emptyset} \lambda_S(v) w_S$ with respect to the basis $W$ defined by (\ref{eq:basis3}). Now, given $v\in G(N)$, we define $D_W(v)=\{S\subseteq N: \lambda_S(v) \neq 0,v=\sum_{S\subseteq N: S\neq \emptyset} \lambda_S(v) w_S\}$ and $d_W(v)=|D_W(v)|$. For $v\in G(N)$ with $d_W(v)\geq 2$, we define the graph $(N, G_v,W)$ where every pair $\{i,j\}$ of players in $N$ forms a link in $G_v$, $i\neq j$ if and only if there exists an $S\in D_W(v)$ with $\{i,j\}\subseteq S$ or $\{i,j\}\cap S=\emptyset$. With an abuse of notation, we denote a link in $G_v$ by the pair $\{i, j\}$ itself.
 A coalition $B$ is connected in $G_v$ if either $|B|=1$ or for every $i,j\in B$, there exist a sequence of players $i_1, i_2,...,i_m$ such that $i_1=i$, $i_m=j$ and $\{i_p,i_{p+1}\}\in G_v$ for all $p\in \{1,2,3,...,m-1\}$. A connected coalition $B$ is a component or maximal connected coalition in $G_v$ if $\{i,j\}\not \in G_v$ whenever $i\in B$ and $j\in N\setminus B$. Two distinct components are disjoint. Thus we have the following proposition.
 
 \begin{proposition}\label{lemma:8}\rm
The graph $G_v$ has at most two components for $d_W(v)\geq 2$. If $G_v$ has two components then their union is $N$. Also $v$ can be written as $v=\lambda_{B_1}(v) w_{B_1}+\lambda_{B_2}(v) w_{B_2}$ where $B_1, B_2$ are two components of $G_v$.   
 \end{proposition}

\begin{proof}
Suppose that $B_1, B_2, B_3$ be three distinct components in $G_v$. Since $d_W(v)\geq 2$ therefore $D_W(v)\neq \emptyset$. Assume without loss of generality there exist a coalition $T\in D(v)$ such that $T\subseteq B_3$. 
Let $i\in B_1$, $j\in B_2$. Then $\{i,j\}\cap B_3=\emptyset$. Since $T\subseteq B_3$ therefore $\{i,j\}\cap T$=$\emptyset$. Therefore $\{i,j\}\in G_v$. Since $B_1$ is a connected component and $i\in B_1, j\in N\setminus B_1$ therefore $\{i,j\}\not \in G_v$. This is a contradiction. Therefore $G_v$ has at most two components. Suppose that $B_1\cup B_2\neq N$. Then for $i\in B_1, j\in B_2, h\in N\setminus B_1\cup B_2$, we have $\{i,h\}\subseteq N\setminus B_2$, $ \{j,h\}\subseteq B_1$. Therefore $\{\{i,h\},\{j,h\}\}\subseteq G_v$. This is again a contradiction. Therefore $N=B_1\cup B_2$.\\
Suppose that $v=\sum\lambda_{S_i}(v) w_{S_i}$. Since $S_1\subseteq N=B_1\cup B_2$. Assume that $S_1\cap B_1\neq \emptyset $. Since all players in $S_1$ are connected therefore $S_1\subseteq B_1$. If there exist a player $j\in B_1\setminus S_1$ then $\{j,t\}\cap S_1=\emptyset$ for all $t\in B_2$. Then $\{j,t\}\in G_v$. This is another contradiction. Hence $S_1=B_1$. \\
Therefore each $S_i$ is equal to one of $B_1,B_2$ and hence $v=\lambda_{B_1}(v) w_{B_1}+\lambda_{B_2}(v)w_{B_2}$.  
\end{proof}
\indent The next characterization theorem of the $k$-SED value based on fairness goes as follows. 
\begin{theorem}\label{them:4}\rm
A solution $\Phi:G(N)\rightarrow \mathbb{R}^n$ satisfies Eff, $k$-NNPP and F if and only if it is equal to the $k$-SED value. 
\end{theorem}
\begin{proof}
Every solution that satisfies Sym and Lin also satisfy F. Therefore $\Phi^{k-\rm{SED}}$ satisfies F. From subsection~\ref{subsection:4.1}, it follows that the $\Phi^{k-\rm{SED}}$ satisfies Eff and $k$-NNPP. \\
Conversely, suppose that $\Phi$ satisfies Eff, $k$-NNPP and F. If $d_W(v)=0 ~\mbox{or}~ 1$ then $v\in \{v_0, \lambda_T(v) e_T, \lambda_S(v) u_S\}$ for some coalitions $T, S$ such that $t< k, s\geq k$. Then by proposition \ref{lemma:7}, $\Phi_i(v)$ is uniquely determined by the $k$-SED value. Now we apply induction on $d_W(v)$.\\
Assume that $\Phi(v')$ is uniquely determined for all $d_W(v')<k$.  Assume also that $d_W(v')\geq 2$. Then $n\geq 2$. Following similar procedure used in the proof of Theorem 2.5 (pg 311) in \cite{brink2}, we can also show that $\Phi_i(v)$ is unique for all $i\in N$. 
\end{proof} 
\indent Similar to Corollary \ref{cor:1}, an immediate consequence to Theorem~\ref{them:4} is the following.
\begin{corollary}\label{cor:2}\rm
A solution $\Phi:G(N)\rightarrow \mathbb{R}^n$ satisfies Eff, NPP and F if and only if $\Phi=\Phi^{ED}$.
\end{corollary} 
\subsubsection*{Logical Independence of the axioms of Theorem~\ref{them:4}}
\noindent\textbf{Dropping Eff}: The value $\Psi^1$ defined by $\Psi^1_i(v)=\dfrac{1}{2^{n-1}}\sum_{S\subseteq N: s\geq k:i\not \in S}\big\{v(S\cup i)-v(S)\big\}$ satisfies F and $k$-{NNPP} but does not satisfy Eff.\\ 
\noindent\textbf{Dropping $k$-NNPP}: The egalitarian rule $\Phi^{ED} :G(N)\rightarrow \mathbb{R}^n$ given by $\Phi^{ED}_i(v)=\dfrac{v(N)}{n}$ for all $i\in N$, satisfies Eff and F but it does not satisfy the $k$-{NNPP} for $k<n$.\\
\noindent\textbf{Dropping F}: The value $\Psi^2$ defined by $\Psi^2_i(v)=\dfrac{\Psi^1_i(v)}{\sum_{j\in N}\Psi^1_j(v)}v(N)$ for all $i\in N$ (if $\sum_{j\in N}\Psi^1_j(v)\neq 0$) satisfies Eff and $k$-{NNPP} but does not satisfy F. 
\begin{remark}
Note that F is equivalent to DM (see \cite{casajus11}) and therefore, we can replace F by DM in Theorem \ref{them:4} and obtain another characterization as follows:
\begin{theorem}\label{them:44}\rm
A solution $\Phi:G(N)\rightarrow \mathbb{R}^n$ satisfies Eff, $k$-NNPP and DM if and only if it is equal to the $k$-SED value. 
\end{theorem}

\end{remark}
\subsection{The Coalition specific incentive for cooperation}
Here, we introduce another axiom: Coalition specific incentive for cooperation(CSIC) that endogenizes the choice of the size $k$ of the $k$-SED value. 
\begin{enumerate}[\textbf{Axiom} 15.] 
	\item Coalition specific incentive for cooperation(CSIC): A value $\Phi: G(N) \rightarrow \mathbb{R}$ satisfies the Coalition specific incentive for cooperation property if there exists an $S \subset N$ such that $\Phi$ has the property $\Phi_i(e_{T})=0,\,\, \textrm{for all}\; T \subset S\; \textrm{with}\;i\in T,$ and $\Phi_i(u_{S \cup T})=0,\,\, \textrm{for all}\; T\subseteq N \setminus S\;\textrm{with}\; i\not \in S \cup T$.
\end{enumerate}

\noindent Axiom 15 implies that for a value, that satisfies CSIC, there always exists a coalition $S$ of $N$ such that the players in $S$ have no incentive to deviate from $S$ to make smaller coalitions and, the remaining players have no incentive to leave any coalition that includes all the players of $S$. Recall from Section \ref{sec:1} that, such an $S$ for example, can represent a small homogeneous community, a tribe or a clan who do not have incentives to further break into smaller groups. Following theorem states that if such an $S$ exists, its size $s$ is exactly the value of $k$ in the corresponding $k$-SED value.

\begin{theorem}
A solution $\Phi: G(N)\rightarrow \mathbb{R}^n$ satisfies { Eff, Sym, Lin,} and {CSIC} if and only if there exists a unique $k\in \{1,...,n\}$ such that $\Phi=\Phi^{k-SED}$.
\end{theorem}
\begin{proof}
Suppose that there exists a $k\in \{1,...,n\}$ such that $\Phi=\Phi^{k-SED}$. By theorem \ref{them:1}, $\Phi^{k-SED}$ satisfies { Eff, Sym,} and {Lin}. Let $S$ be a coalition of size $k$. Then by the definition of  $\Phi^{k-SED}$, we have $\Phi_i(e_{T})=0\,\, \textrm{for all}\; T \subset S\; \textrm{with}\;i\in T,$ and $\Phi_i(u_{S \cup T})=0\,\, \textrm{for all}\; T\subseteq N \setminus S\;\textrm{with}\; i\not \in S \cup T$. Thus, $\Phi^{k-SED}$ satisfies {CSIC}.  \\
Conversely, let $\Phi$ satisfy {Eff, Sym, Lin,} and {CSIC}. Observe that, if $\Phi: G(N) \rightarrow \mathbb{R}$ satisfies the CSIC then there is an $S\subseteq N$ for which $\Phi$ satisfies the given properties. Take $k=s$. Then, there exists a chain of coalitions $\emptyset=S_0\subsetneq S_1 \subsetneq ...\subsetneq S_{n-1}\subsetneq S_n=N$ with $s_j=j$ for $j\in \{0,1,...,n\}$ and $S= S_k$ such that $\Phi$ satisfies the following conditions:
\begin{enumerate}[~~~~(a)]
 \item $\Phi_i(e_{S_j})=0,\,\, \forall i\in S_j,\,\, j\in\{1,... k-1\}$ and 
\item $\Phi_i(u_{S_j})=0,\,\, \forall i\not \in S_j,\,\, j\in\{k,...,n\}$.
\end{enumerate}
Since $\Phi$ is an ESL value, therefore, by proposition \ref{prop:esl1}, there exists a unique collection of real constants $\{b^{\Phi}_i: i\in \{1,2,3,...,n\}\}$ with $b^{\Phi}_0=0$, $b^{\Phi}_n=1$ such that 
\begin{equation}\label{eq:esl3}
	\Phi_i(v)=\sum_{S\subseteq N\setminus i} \dfrac{s!(n-s-1)!}{n!}\bigg\{b^{\Phi}_{s+1}v(S\cup i)-b^{\Phi}_sv(S)\bigg\}.
\end{equation}
Recall from Remark~\ref{rem:2} that for any $k\in \{1,...,n\}$, $\Phi$ given by (\ref{eq:esl3}) coincides with $\Phi^{k-SED}$ only if $b^\Phi_j=0$ for $j<k$ and $b_j^\Phi=1$ for $j\geq k$.
In view of the observation made in the beginning, there exists a sequence of coalitions $\emptyset=S_0\subsetneq S_1 \subsetneq ...\subsetneq S_{n-1}\subsetneq S_n=N$ with $s_j=j\in \{0,1,...,n\}$, and a $k\in\{1,...,n\}$ such that $\Phi_i(e_{S_j})=0\,\, \forall i\in S_j$, $j\in \{1,...,k-1\}$ and $\Phi_i(u_{S_j})=0\,\, \forall i\not \in S_j,\,\, j\in \{k,...,n\}$. Since $\Phi_i(e_{S_j})=\dfrac{(n-s_j)!(s_j-1)!}{n!}b^{\Phi}_{s_j}$ for $i \in S_j$ therefore, $b^{\Phi}_{s_j}=b^\Phi_j=0$ for $j<k$. If $k=n$ then $b^{\Phi}_{j}=0$ for all $j<n$. Therefore, $\Phi=\Phi^{n-SED}$. Assume that $1\leq k<n$ and $i \not \in S_{n-1}$. Then  $\Phi_i(u_{S_{n-1}})=\frac{1}{n}(b^{\Phi}_n-b^{\Phi}_{n-1})$. Since $n-1\geq k$, therefore, $\Phi_i(u_{S_{n-1}})=0$. Thus, $b^{\Phi}_n=b^{\Phi}_{n-1}=1$. If $n-2\geq k$, then $\Phi_i(u_{S_{n-2}})=\frac{1}{n}(b^{\Phi}_{n-1}-b^{\Phi}_{n-2})$ for all $i\not \in S_{n-2}$. Thus, $b^{\Phi}_{n-1}=b^{\Phi}_{n-2}$. Continuing in this way, we have $1=b^{\Phi}_n=b^{\Phi}_{n-1}=b^{\Phi}_{n-2}=...=b^{\Phi}_{k}$. Thus, $\Phi=\Phi^{k-SED}$.\\ 
Suppose that for another coalition $S'$, $\Phi$ satisfies {Eff, Lin, Sym} and {CSIC}. Then, under the above observation, we obtain another sequence of coalitions $\emptyset=T_0\subsetneq T_1 \subsetneq ...\subsetneq T_{n-1}\subsetneq T_n=N$, with $S'= T_{k'}$ for some $k'$. But, then $k'= s'$ and we have $b^{\Phi}_0=b^{\Phi}_1=b^{\Phi}_3=...=b^{\Phi}_{k'-1}=0$ and $b^{\Phi}_{k'}=b^{\Phi}_{k'+1}=...=b^{\Phi}_n=1$. Thus, $k=k'$ and therefore, $k$ is unique.
\end{proof}

\section{Comparison with existing solutions}\label{sec:5}
In~\cite{yokote17}, it is conjectured that there exists a large class of linear values which can be characterized by monotonicity. The $k$-SED value being a linear solution characterized by $k$-PMon therefore, belongs to this large class of values. 

Our value is also closely related to the solidarity value $\mathbf{Sol}_N$ proposed recently in \cite{beal}. For an integer $p$ in $\{0,1,2,...,n-1\}$, the payoff to player $i\in N$ given by $Sol^p$ due to \cite{beal} has the following form.
\begin{equation}\label{eq: bealsol1}
	Sol_i^p(v)=\sum_{{S\subseteq N:\\ i\in S, s\leq p}} \dfrac{(n-s)!(s-1)!}{n!}\bigg (v(S)-v(S\setminus i)\bigg )+\sum_{S\subseteq N: i \not \in S, s=p} \dfrac{(n-s-1)!s!}{n!}\bigg (v(N)-v(S)\bigg ).
\end{equation}
If $p$ is drawn from $\{0,1,2,...,n-1\}$ according to the probability distribution $\beta=\{\beta_p:p\in \{0,1,2,...,n-1\}\}$ then the solidarity value $\mathbf{Sol}_N$ due to \cite{beal} induced by the probability distribution $\beta$ is defined as follows:
\begin{equation}
\mathbf{Sol}_N(v)=\sum_{p=0}^{n-1} \beta_p Sol^p(v)
\end{equation}
It is easy to verify that if $k=n-p$, for $p\in \{0,1,2,...,n-1\}$, then $\Phi^{k-\rm{SED}}(v)=Sol^p(v^*)$ where $v^*$ given by $v^*(S)=v(N)-v(N\setminus S)\;\;\forall S\subseteq N$, is the dual game of $v$.  This is because $\Phi^{k-\rm{SED}}$ and $\mathbf{Sol}_N$ represent two opposite social situations. In case of $\mathbf{Sol}_N$ which is indeed the expected payoff vector of the average payoff vectors $Sol^k$, $k=0,1,...n-1$, each player entering at position $\pi(i) \leq k$ obtains her contribution $v(P(\pi, i))-v(P(\pi, i)\setminus i)$ upon entering while each player entering at position $\pi(i)>k$ obtains an equal share of the remaining worth $v(N)- v(P(\pi, \pi^{-1}(k)))$. The $k$-SED value $\Phi^{k-\rm{SED}}$ on the other hand, awards equal share to each player entering at position $\pi(i) \leq k$ and her marginal contributions when she enters at the position $\pi(i)>k$. 

\noindent One of the key axioms in both \cite{beal} and our model to characterize the values $Sol^p$ and $\Phi^{k-\rm{SED}}$ involves a type of null player. It is the $p$-null player for $Sol^p$ and the $k$-nullifying null player for $\Phi^{k-\rm{SED}}$. \\
Given $p\in \{1,2,3,...,n-1\}$, $v\in G(N)$, a player $i\in N$ is called $p$-null player in $v$ if 
\begin{equation*}
\forall S \subseteq N, i\in S, s\leq p, v(S)=v(S\setminus i) ~~~\text{and}~~~ \forall S \subseteq N, i\not \in S, s=p, v(N)=v(S).
\end{equation*}  
A value $\Phi$ satisfies $p$-null player axiom if for each $p$-null player in $v$, it holds that $\Phi_i(v)=0$. The two players, the $k$-nullifying null player and the $p$-null player are similar, but they build on two completely different social narratives. Unlike the $k$-nullifying null player, the $p$-null player is non-productive in all coalitions till they reach a size $p$, and the worth of all coalitions of size $p$ where she is not a member is equal to the worth of the grand coalition. This axiom may possibly be considered somewhat  restrictive and demanding and is particularly specific to the requirement of the formulation of the solidarity value $\mathbf{Sol}_N$. It is not clear to what extent one can argue that it represents some social criterion. It is a matter of further study that, why and how a player without being a member of a coalition of size $p$ can influence that coalition to generate the same worth as that of the grand coalition! The $k$-nullifying null player on the other hand divides the class of coalitions into two groups, in one group it acts as a nullifying player and in the other, as a null player. When $k = 1$ it is the null player and for $k=n$ it is the nullifying player. Note that the standard characterization of the Shapley value and the ED requires the null player property\cite{shapley} and the nullifying player property \cite{brink2007} respectively. Thus it also supports our intuition that the $k$-nullifying null player inherits characteristics from both null and nullifying types of players. 

\noindent We conclude this section with the following few observations from \cite{beal}. For details, refer to 
Propositions $6$, $8$ and $11$ in \cite{beal}.

\begin{proposition}\label{prop:4}(Proposition 6 of \cite{beal}, p. 72). Fix any $p=\{0,...,n-1\}$. If $p=0$ then $Sol^0=ED$ and if $p=n-1$ then $Sol^{n-1}=\Phi^{Sh}$. 
\end{proposition}

\begin{proposition}\label{prop:5}(Proposition 8 of \cite{beal}, p. 73). A value $\Phi$ on $G(N)$ belongs to $\textbf{Sol}_N$ if and only if it can be represented by 
\begin{equation*}
\Phi_i(v)=\sum_{S\subseteq N\setminus i} \dfrac{s!(n-s-1)!}{n!}\bigg\{b^\Phi_{s+1} v(S\cup i)-b^\Phi_s v(S)\bigg\}
\end{equation*}
with constants $B^\Phi=\{b_s^\Phi:s\in \{0,1,2,...,n\} \}$ such that 
\begin{equation*}
b_0^\Phi=0, b_n^\Phi=1, ~~\mathtt{ and }~~ \forall s\in \{1,2,...,n-1\},~~ 1\geq b_1^\Phi \geq b_2^\Phi \geq ... \geq b_{n-1}^\Phi \geq 0.
\end{equation*}
Furthermore, $\Phi=Sol^\beta$, where $\beta=\{\beta_s: s\in \{0,1,2,...,n-1\}\}$ is obtained from the transformation $B^\Phi \mapsto \beta$ such that 
\begin{equation*}
\beta_0=1-b_1^\Phi, \beta_{n-1}=b_{n-1}^\Phi ~~~\text{and}~~~ \forall s \in \{1,2,...,n-2\} , \beta_s=b_s^\Phi-b_{s+1}^\Phi.
\end{equation*}
\end{proposition}
\begin{proposition}\label{prop:6}(Proposition 11 of \cite{beal}, p. $80$) A value $\Phi$ on $G(N)$ is equal to $Sol^p$ for $p\in \{1,2,...,n-1\}$ if and only if it satisfies Eff, Equal treatment of equals, Additivity and the $p$-null player axiom.
\end{proposition}
\noindent Proposition \ref{prop:4}~ above, shows that $Sol^0 = \Phi^{ED}$ and $Sol^{n-1} = \Phi^{Sh}$ and Proposition \ref{prop:6} characterizes $Sol^p$, $p \in \{1,2,...,n-1\}$ using Eff, equal treatment of equals (Sym in our terminology), ADD and the $p$-null player axiom. However, in view of Proposition \ref{prop:4}, the axioms in~ Proposition \ref{prop:6} cannot characterize the ED as $Sol^p$ is equal to the ED, only when $p=0$. Therefore, the characterization due to Proposition \ref{prop:6} does not seem to be a complete characterization from the Shapley value to the ED. On the other hand, all the four characterizations we have proposed here, completely characterize the range of the $k$-SED values starting from $k=1$ to $k=n$ including the Shapley value and the ED at the two extremes. 

\section{The $\alpha$-SED value and implementation}\label{sec:6}
Generalizing the $k$-SED value to include all possible values of $k \in N$ in the line of \cite{beal}, we define the $\alpha$-SED value as follows. Assume that the integer $k$ is drawn from $N$ according to the probability distribution $\alpha= (\alpha_k:k\in N)$. Then the $\alpha$-SED value $\Phi^{\alpha-SED}$ induced by the probability distribution $\alpha$ is defined for $v\in G(N)$ as the expected payoff given by  
\begin{equation}\label{Eq:SED}
\Phi^{\alpha-SED}(v)=\sum_{k=1}^{n} \alpha_k\Phi^{k-SED}(v).
\end{equation}
Thus, the $\alpha$-SED value computes the expected payoff of each player under the probability distribution 
$\alpha$. 
\begin{remark}\label{rem:1}
	\begin{enumerate}[~~(a)]
		\item If $\alpha_1=1$, $\alpha_k=0$ for $k\geq 2$ then $\Phi^{\alpha-SED}= \Phi^{Sh}$.
		\item If $\alpha_n=1$, $\alpha_k=0$ for $1\leq k\leq n-1$ then $\Phi^{\alpha-SED}=\Phi^{ED}$.
		\end{enumerate}
\end{remark}

Now we propose a mechanism that implements the $k$-SED and the $\alpha$-SED values for zero-monotonic games in Subgame Perfect Nash Equilibrium (SPNE). This mechanism is an adaptation of the mechanism proposed in  \cite{beal}. The only difference in the proposed mechanism from the mechanism of \cite{beal} is in the take-it-or-leave-it part of \textbf{stage 3} of the \textbf{Mechanism (B)}. In the step 4 of stage 3 of Mechanism B in \cite{beal}, all the players in position higher than (and including) $p+1$ leave with null payoff when the offer proposed by player in position $p+1$ is not accepted by consensus. In our mechanism it will be the set of players in position lower than (and including) $p+1$ (In our formulation, it is $k+1$). For the sake of completeness we will define the modified mechanism below and provide the proposition without proof. 

\noindent \textbf{Mechanism (M)}

\noindent Consider any TU-game $v\in G(N)$ and a probability distribution $\alpha = {(\alpha_k)}_{k=1}^{n}$ on $N$ and define $A= \{i\in N|\alpha_i \ne 0\}$ the support of $\alpha$.

\noindent \textbf{Stage 1}: Each player $i\in N$ makes bids $h_k^i \in \mathbb{R}$, one for each position $k \in A$, under the following constraint:

\begin{equation}
\sum_{k\in A} \alpha_k h_k^i =0.
\end{equation}
\noindent For each position $k \in A$, define the aggregate bid $H_k$ as:
\begin{equation}
H_k =\sum_{i\in N} h_k^i.
\end{equation}

\noindent Denote by $\Omega_A$ the subset of positions with the highest aggregate bid.

\noindent \textbf{Stage 2}: Each player $i \in N$ makes bids $h^i_\pi\in \mathbb{R}$, one for each permutation $\pi\in\Pi(N)$, under the constraint:

\begin{equation}\label{eq:ass1}
\sum_{\pi\in \Pi(N)} \frac{1}{n!}h^i_\pi =0.
\end{equation}

The condition given by Eq.(\ref{eq:ass1}) suggests that the designer gives each permutation $\pi$ equal weights, namely $\frac{1}{n!}$. For each permutation $\pi\in\Pi(N)$, the aggregate bid, defined in a similar way as in \textbf{Stage 1}, is denoted by $H_{\pi}$. Finally, denote by $\Omega_{\Pi(N)}$ the subset of permutations with the highest aggregate bid.\\

\noindent \textbf{Stage 3}: Pick at random any $k\in\Omega_A$ and then any $\pi\in \Omega_{\Pi(N)}$. Together, position $k$ and permutation $\pi$ induce a sequential bargaining game $G_{k,\pi}$ whose payoffs are denoted by ${(g^i_{k,\pi})}_{i \in N}$. This bargaining game contains the following steps:
\begin{enumerate}[~~~(i)]
\item Player $i \in N$ in position $\pi(i) = k+1$ proposes an offer $x^i_j \in \mathbb{R}$ to each other $j \in N\setminus i$.
\item The players other than player $i$, sequentially, either accept or reject the offer. If at least one player rejects it, then the offer is rejected. Otherwise the offer is accepted.
\item If the offer is accepted, then the payoffs are given by:
\begin{equation}
g^i_{k,\pi} = v(N) - \sum_{j \in N\setminus i} x^i_j, \mbox{ and } \forall j \in N\setminus i, \mbox{  }g^j_{k,\pi} = x^i_j.
\end{equation}
\item If the offer is rejected, then each player $j$ in position $\pi (j) \leq k+1$ leaves the bargaining procedure with a null payoff, i.e. $g^j_{k,\pi} = 0$, while all the players $j$ in positions $\pi (j) \geq k+2$ proceed to the next round to bargain over $v(P^{\pi^{-1}(k)}(\pi))$.
\item The new proposer is player $i$ in position $\pi (i) = k+2$. Player $i$ makes an offer $x^i_j \in \mathbb{R}$ to each other player $j$ such that $\pi (j) \geq k+3$. If the offer is unanimously accepted by all the players $j$ in position $\pi (j) \geq k+3$, then the payoffs are as follows:
\begin {equation}
g^i_{k,\pi} = v(P^i(\pi)) - \sum_{\pi (j)>\pi(i)} x^i_j, \mbox{ and } \forall j : \pi (j)>\pi(i), \mbox{  }g^j_{k,\pi} = x^i_j.
\end{equation}
If the offer is rejected, then player $i$ in position $\pi (i) = k+2$ leaves the bargaining procedure with a null payoff. Then, stage (iv) is repeated among the players $j$ in position $\pi (j) \geq k+3$, where the new proposer is player in position $k+3$. Stage (v) is repeated until a proposal is accepted. In case the bargaining procedure reaches the situation where the only active player $i$ is such that $\pi (i) = n$, then his or her payoff in $G_{k,\pi}$ is equal to $v(i)$.
\end{enumerate}

\noindent \textbf{Stage 4}: Rewards $(z^i_{k,\pi})_{i\in N}$ resulting from \textbf{Stage 1, 2} and \textbf{3} in $G_{k,\pi}$ are defined as:

\begin{equation}
\mbox{  } z^i_{k,\pi} = g^i_{k,\pi} - h^i_k - h^i_\pi + \frac{H_k + H_\pi}{n}, \forall i \in N.
\end{equation}

\noindent That is, each player pays his or her bids, receives an equal share of the aggregate bids $H_k$ and $H_\pi$ plus the payoff resulting from the bargaining procedure $G_{k,\pi}$. Finally, since $k$ and $\pi$ are chosen randomly in $\Omega_A$ and $\Omega_{\Pi(N)}$, the expected payoff of each player playing \textbf{Mechanism (M)} is given by

\begin{equation}
\forall i \in N, \mbox{  } m_i = \frac{\sum_{k\in \Omega_A} \sum_{\pi \in \Omega_{\Pi(N)}} z^i_{k,\pi}}{|\Omega_A| \times |\Omega_{\Pi(N)}|}
\end{equation}

\begin{proposition}\label{prop:implement}\rm
Consider any zero-monotonic TU game $v\in G_N$ and a probability distribution $\alpha$ with support $A \subseteq \{1, \cdots , n\}$. Then, \textbf{Mechanism (M)} implements the $\alpha$-SED value in SPNE.
\end{proposition}

\begin{proof}
The proof follows the steps of the proof in \cite{beal} closely adjusting for the order of the agents in take-it-or-leave-it step. Therefore we omit the proof.
\end{proof}

\section{Conclusion}\label{sec:7}
We started with the proposal of a new value for TU games that exhibits the characteristics of the Equal Division rule in small coalitions and the Shapley value in sufficiently large coalitions. The procedure for obtaining this value is driven by the assumption that players in small groups share their collective resource in an egalitarian manner, but become more competitive with the increase of the size of the group. In large groups, where people are less likely to be altruistic, it is the marginal productivity of each member that controls a fair distribution of their resources. This value clearly reflects such a social phenomenon. Our model generates a whole range of values that includes the Shapley value and the ED at its two extremes. There are possibilities to explore alternative characterizations of the proposed value. We keep this for our future research.
 
\subsection*{Acknowledgement}
This research was funded by UKIERI[184-15/2017(IC)].


\begin{thebibliography}{20}

\bibitem{beal}
B\'eal, S., R\'emila, E. and Solal, P. (2017) Axiomatization and implementation of a class of solidarity values for TU-games, Theory and Decision, 83, 61--94. 

\bibitem{beal2}
B\'eal, S., R\'emila, E. and Solal, P. (2017) A strategic implementation of the sequential equal surplus division rule for digraph cooperative games, Annals of Operations Research, 253, 43--59. https://doi.org/10.1007/s10479-016-2290-5.

\bibitem{tijs}
Branzei, R., Dimitrov, D. and Tijs, S.(2008) Models in Cooperative Game Theory: Crisp, Fuzzy and Multichoice Games, Springer, Berlin Heidelberg.

\bibitem{casajus14}
Casajus, A. and Huettner, F. (2014) Weakly monotonic solutions for cooperative games, Journal of Economic Theory, 154, 162--172.

\bibitem{casajus12}
Casajus, A. and Huettner, F. (2013) Null players, solidarity, and the egalitarian Shapley values, Journal of Mathematical Economics, 49, 58--61.

\bibitem{casajus11}
Casajus, A. (2011) Differential marginality, van den Brink fairness, and the Shapley value, Theory and Decision, 71, 163--174.

\bibitem{chun}
Chun, Y. (1989) A New Axiomatization of the Shapley values, Games and Economic Behavior, 1, 119--130.

\bibitem{driessen1}
Driessen, T. (1988) Cooperative Games, Solutions and Applications, Kluwer Academic Publishers, Springer Netherlands.

\bibitem{driessen}
Driessen, T., and Radzik, T. (2002) A weighted pseudo-potential approach to values for TU games, International Transactions in Operational Research, 9, 1--18.

\bibitem{driessen2003}
Driessen, T., and Radzik, T., (2003) Extensions of Hart and Mas-Colell’s consistency to efficient, linear, and symmetric values for TU-games. In: Petrosyan, L.A., Yeung, D.W.K. (Eds.), ICM Millennium Lectures on Games. Springer-Verlag, Heidelberg, Germany, 147--166. 

\bibitem{joosten}
Joosten, R., (1996) Dynamics, equilibria and values dissertation. Maastricht University.

\bibitem{radzik}
 Nowak, A. S., and Radzik, T., (1994) A solidarity value for n-person transferable utility games, International
Journal of Game Theory, 23, 43--48.

\bibitem{radzik13}
Radzik, T., and  Driessen, T., (2013) On a family of values for TU-games generalizing the Shapley value, Mathematical Social Sciences, 65, 105--111.

\bibitem{ESLvalue}
Radzik, T., Driessen, T., (2016) Modeling values for TU-games using generalized versions of consistency, standardness, and the null player property, Mathematical Methods of Operations Research, 83, 179--205. 

\bibitem{shapley}
Shapley, L. S. (1953) A value for n-person games, in Kuhn, H. and Tucker, A.W. (eds.), Contribution to the Theory of games II, Princeton, New Jersey, Princeton University Press, 307--317.

\bibitem{sprumont}
Sprumont, Y. (1990) Population monotonic allocation schemes for cooperative games with transferable utility, Games and Economic Behavior 2, 378--394.

\bibitem{brink2}
van den Brink, R. (2001) An axiomatization of the Shapley value using a fairness property, International Journal of Game Theory 30, 309--319.

\bibitem{brink2007}
van den Brink, R. (2007) Null or nullifying players: the difference between the Shapley value and equal division solutions, Journal of Economic Theory, 136, 767--775.

\bibitem{brink2013}
van den Brink, R.,  Funaki, Y., and Ju, Y., (2013) Reconciling marginalism with egalitarianism: consistency, monotonicity, and implementation of egalitarian Shapley values, Social Choice and Welfare, 40, 693--714.

\bibitem{brink_funaki}
van den Brink, R. and Funaki, Y., (2015) Implementation and axiomatization of discounted Shapley values, Social Choice and Welfare, 2,45, 329--344.

\bibitem{yokote17}
Yokote, K. and Funaki, Y., (2017) Monotonicity implies linearity: characterizations of convex combinations of solutions to cooperative games, Social Choice and Welfare, 49, 171--203.

\bibitem{young}
Young, H. P. (1985) Monotonic solutions of cooperative games, International Journal of Game Theory, 14, 65--72.

\end{thebibliography}
\end{document}